# Overcoming the tradeoff between confinement and focal distance using virtual ultrasonic optical waveguides

Matteo Giuseppe Scopelliti[1], Hengji Huang[1], Adithya Pediredla[2], Srinivasa Narasimhan[2], Ioannis Gkioulekas[2], Maysamreza Chamanzar[1]

[1]Electrical and Computer Engineering Department, Carnegie Mellon University, Pittsburgh PA, 15213 USA.
[2]Robotics Institute, Carnegie Mellon University, Pittsburgh PA 15213, USA.

**Abstract**

Conventional optical lenses have been used to focus light from outside without disturbing the medium. The focused spot size is proportional to the focal distance in a conventional lens, resulting in a tradeoff between depth of penetration in the target medium and spatial resolution. We have shown that virtual ultrasonically sculpted gradient-index (GRIN) optical waveguides can be formed in the target medium to guide and steer light without disturbing the medium. Here, we demonstrate that such virtual waveguides can relay an externally focused beam of light through the medium beyond the focal distance of an external physical lens to extend the penetration depth without compromising the spot size. Moreover, the spot size can be tuned by reconfiguring the virtual waveguide. We show that these virtual GRIN waveguides can be formed in transparent as well as turbid media to enhance the confinement and contrast ratio of the focused beam of light at the target location. This method can be extended to realize complex optical systems of external physical lenses and in situ virtual waveguides to extend the reach and flexibility of optical methods.

**Introduction**

Light-matter interaction has been used in different applications, ranging from biological imaging and manipulation to metrology, material processing, and machine vision[1–7]. Through the interaction of light with matter, a signature of the medium may affect light, which can be used for sensing, detection, and imaging. Moreover, light can affect the medium when it is concentrated to a high enough intensity at specific locations within the medium. Optical manipulation has been used in a wide range of applications such as optogenetic stimulation of biological events, photothermal therapy of cancer tumors, 3D printing, machining, and material processing. A key advantage of using light, whether for probing or manipulation, is that it can penetrate through materials non-invasively at appropriate wavelengths. Different optical components such as lenses, spatial light modulators, and waveguides have been used to shape the trajectory of light. Tunable external optical components such as optical modulators, tunable lenses, and gratings have also been realized based on electro-optic or acousto-optic effects to reconfigure the pattern of light before it is launched into the target medium[8–11]. The ability to manipulate the trajectory of light within the target medium will extend the power and flexibility of optical methods. Invasive insertion of traditional optical components such as lenses or waveguides into the medium defeats the purpose of using light as a non-invasive modality for interaction with the medium, especially in the case of non-destructive testing of materials or imaging and stimulation of biological tissue. We have recently shown that ultrasound waves can be used to guide and pattern the trajectory of light by locally changing the refractive index of the medium. Using this technique, we have demonstrated the possibility of forming in situ virtual gradient-index (GRIN) waveguides, virtual relay lenses, and spatial light modulators non-invasively[12–14]. Since ultrasound in the proper frequency range can propagate deep with minimal attenuation in the medium, virtual optical components can be realized within the target medium to manipulate the trajectory of light without inserting any physical devices disturbing the medium. We have shown that these virtual optical components can be formed in transparent as well as scattering media such as biological tissue[12]. In this method, the virtual optical component can be reconfigured by simply changing the pattern of ultrasound waves from outside the medium. A nonlinear photoacoustic wave implementation of this idea has also been recently demonstrated for light guiding to deep biological tissue sites[15]. Transversal ultrasound guiding of light deep into scattering media has also been successfully demonstrated using this technique[16].

In this paper, we demonstrate that the virtual ultrasonically sculpted GRIN waveguides can be employed to relay an externally focused beam of light through the medium, non-invasively, without compromising the spot size. Our simulation results suggest that by sculpting the appropriate refractive index pattern within the target transparent medium, light focused by an external lens can be relayed through multiple pitch lengths of a virtual GRIN waveguide while maintaining or even decreasing the focal spot size. Therefore, using this cascade system of an external physical lens and the virtual GRIN waveguide, the tradeoff between the focal distance and the focal spot size of the external lens can be overcome. Moreover, by translating and reconfiguring the pattern of ultrasonic waves, the beam of light can be confined at different locations within the medium, and the confined beam spot size can be tuned. We show that virtual GRIN waveguides can also be formed in scattering media to relay an externally focused beam of light through the medium. The overall contrast between the intensity of light at the focal point and the surrounding local background within the scattering medium is enhanced using this technique instead of an external physical lens with a similar focal length, suggesting that the distribution of both ballistic and scattered photons is affected by the virtual waveguide. These results inspire the tantalizing notion of designing complex optical systems of multiple virtual optical elements sculpted within the target medium in tandem with external optical components to enable unprecedented control over the trajectory of light within the target medium in a non-invasive way.

**Extending the focal distance without compromising the spot size using ultrasonically sculpted virtual GRIN optical waveguides**

Imagine light generated by an external source is supposed to be confined and focused to a point $P$ at a depth $d$ inside a medium with a negligible level of scattering (Fig. 1a). Since the light source is outside the medium, an external lens can be used to confine and focus light onto the target location inside the medium. To achieve the smallest possible focal spot using a lens, the input light must be collimated (Fig. 1b). Given the input beam diameter $D$ and the focal distance $f$, which should be slightly larger than $d$, i.e., $f = d+\Delta L$ so that the external lens can be placed just outside the medium, the focal spot size, characterized by the beam waist at the focal plane assuming negligible aberrations, can be obtained as

$$2w = \frac{4\lambda}{\pi}\frac{f}{D}, \tag{Eq.1}$$

where $2w$ is the focused beam waist, and $\lambda$ is the wavelength of light[17]. The focal spot size is proportional to the focal distance, and therefore, the spatial resolution of the focused beam of light decreases with the depth of penetration. To overcome this limitation and achieve a smaller focal spot and, hence, a higher spatial resolution at depth, an external lens with a shorter focal distance can be used to focus light shallower into the medium and, then a second lens or waveguide can be used inside the medium to relay the tightly focused beam of light to the target point. However, inserting a second lens or waveguide into the medium will disturb the medium and is invasive. Here, we show that a virtual GRIN waveguide can be formed using ultrasound to relay the externally focused beam of light to the target location $P$ without physically disturbing the medium (Fig. 1c).

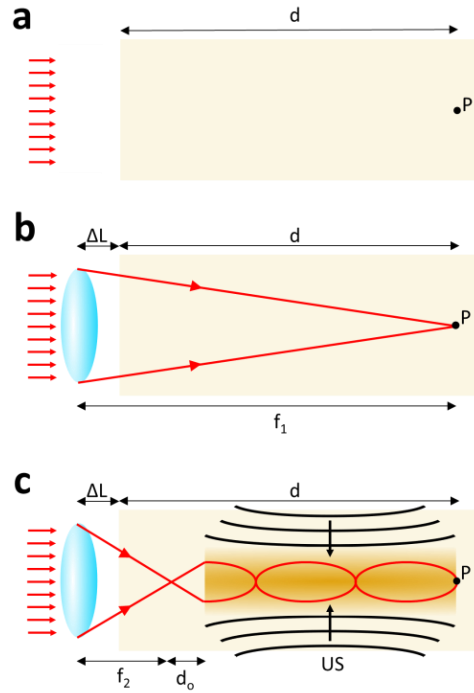

**FIG. 1** a) Schematic illustration of a collimated beam of light from an external source and the target point *P* inside the medium. b) An external lens with a long focal distance ($f_1$) is used to focus light to the target location *P* at a depth *d* inside the medium. c) A cascade optical system of an external lens with a short focal distance ($f_2$) and an ultrasonically defined virtual GRIN waveguide is used to relay the focused spot through the medium to the same point *P*.

We used a commercial optical design software (OpticStudio 20.1, ZEMAX LLC) to perform ray tracing analysis and also calculate the spot size using the Physical Optics Propagation module. First, we simulated an external lens with a long focal distance (LF) of $f_1 = 50$ mm (in air) to directly focus the input collimated beam of light through a non-scattering medium (water) with a refractive index of $n = 1.333$. Taking into account that the lens has to be placed outside the medium at $\Delta L = 11.27$ mm and also the elongation of the focal distance in water (due to the higher refractive index of water compared to air), the beam was focused at a depth of $d = 47.22$ mm into the medium. The axial ray paths and the focused beam profile at the focal plane are shown in Fig. 2a. A radial cross-section of the focused beam is shown in Fig. 2b. The focused beam size, defined as the full width at half maximum (FWHM), was measured as FWHM = 12.05 µm. Next, a cascade optical system of a short focal distance (SF) external lens with $f_2 = 25$ mm (in air) in tandem with an in situ GRIN waveguide was simulated. Such GRIN waveguide was assumed to possess a cylindrically-symmetric parabolic refractive index profile along the radial direction with a maximum refractive index contrast of $\Delta n = 6\times10^{-4}$. The simulation results in Fig. 2c show the ray paths along the axial direction in the cascade optical system, where the beam of light focused by the external lens is relayed along the axial direction by the virtual GRIN waveguide to the same depth $d = 47.22$ mm, for which the distance between the focal point of the external lens and the virtual GRIN waveguide, i.e., $d_o$ had to be $d_o = 13.47$ mm. A radial cross-section of the focused beam at the output plane is shown in Fig. 2d, with the beam size of FWHM= 3.76 µm, which is much smaller than the beam size focused by the single long focal distance (LF) external lens. The focused beam size can be adjusted by changing the distance $d_o$ or the parameters of ultrasound, such as its frequency and amplitude.

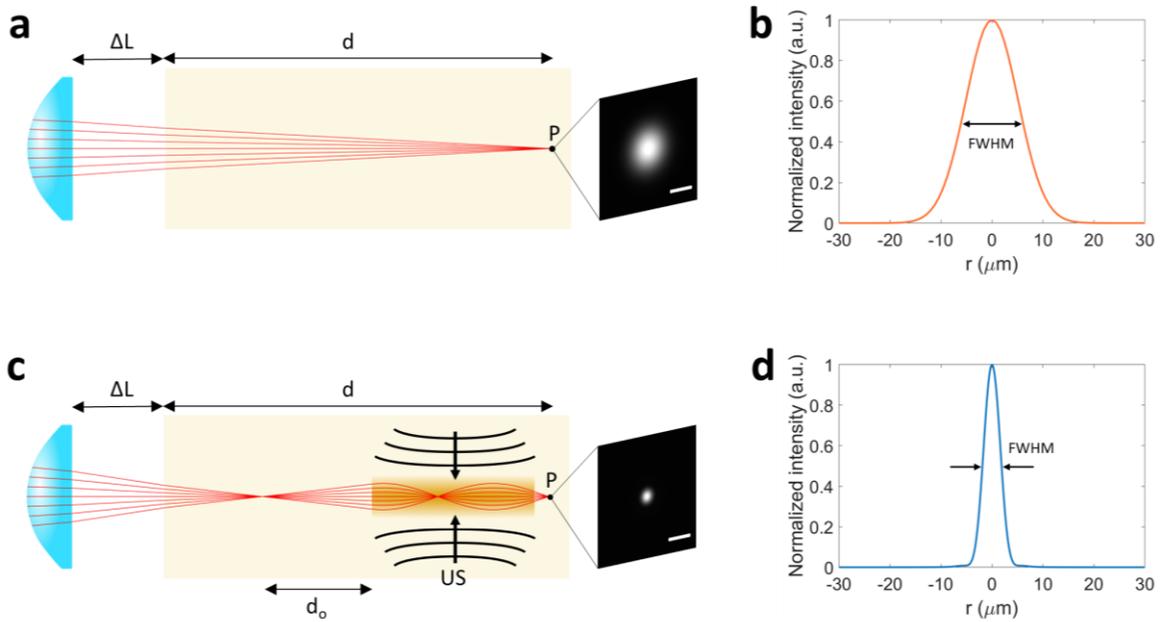

**FIG. 2** a) Optical ray tracing simulation results showing the axial beam paths through the medium and the spot size at the focal plane of a long focal distance lens. b) The radial cross-section of the focused beam of light at point *P*. The diameter of the focal spot is FWHM= 12.05 μm. c) The axial beam paths through the medium and the spot size at the focal plane of the cascade optical system of an external lens and the virtual GRIN waveguide. d) The radial cross-section of the focused beam of light at point *P* using the cascade system. The diameter of the focal spot is FWHM= 3.76 μm. In these simulations, the medium is assumed to be water, $d = 47.22$ mm, $\Delta L = 11.27$ mm, and $d_o = 13.47$ mm. The scale bar is 10 μm.

**Results**

We performed experiments to demonstrate the concept of non-invasive relaying of an externally focused beam of light using the virtual GRIN waveguide both in a transparent and also in a scattering medium. First, we performed an experiment using an external aspherical lens (49102, Edmund Optics Inc.) with a short focal distance (SF) of $f_2 = 25$ mm (in air) cascaded with a virtual ultrasonically sculpted GRIN waveguide to relay the focused beam of light to a depth of $d = 47.1$ mm into deionized (DI) water as a transparent medium. The lens was placed at a distance $\Delta L = 11.27$ mm outside a container made of 3 mm thick acrylic filled with DI water (Fig. 3a). Light at the wavelength of $\lambda = 640$ nm from a fiber-coupled laser (OBIS LX 640nm, Coherent Inc.) was collimated using an adjustable fiber collimator (CFC-2X-A, Thorlabs, Inc.) before impinging on the lens. As shown in Fig. 3a, the focused beam of light was imaged in the transmission mode using a microscope composed of a zoom lens (VZM 600i, Edmund Optics Inc.) directly attached to a CCD camera (BFLY-U3-50H5M-C, FLIR Systems). The zoom lens was capped with a clear optical window (WG11050-A, Thorlabs, Inc.), which was immersed in DI water. To form the virtual GRIN waveguide in DI water, we used a cylindrical piezoelectric transducer made of PZT Type II, with a thickness = 1.5 mm (American Piezo Ceramics, Inc.) driven by a 30 V signal at 1.512 MHz. The relayed confined optical beam is shown in Fig. 3b. We also used an external aspherical lens (33-945, Edmund Optics Inc.) with a longer focal distance (LF) of $f_1 = 50$ mm (in air) to confine light to the same depth of $d = 47.1$ mm into the medium. The focused beam of light at the output plane is shown in Fig. 3c, where we can observe a much larger spot size compared with that of the cascade system. To quantify the difference, the radial cross-sections of the confined beam using the single external lens (i.e., LF) and that of the cascade system of the external lens and the virtual GRIN waveguide (i.e., SF+US) are plotted in Fig. 3d, where we can see that the spot size in the case of a single external lens (FWHM = 53 µm) is larger than the spot size of the shorter focal distance external lens cascaded with the virtual GRIN waveguide (FWHM = 20 µm). Moreover, the peak intensity of the focused beam using the cascade system is 6.98 times higher than the peak intensity of the beam focused by the single external lens, demonstrating that the tradeoff between the focal distance and the diffraction-limited spot size (Eq. 1) in conventional lenses is overcome using the cascade optical system. In other words, we can benefit from the strong confinement of the short focal distance external lens to achieve a tighter focus, while non-invasively relaying the focal spot through the medium using the ultrasonically defined virtual GRIN waveguide. This is a unique advantage of using ultrasonically defined virtual GRIN waveguides that enables high-resolution optical access to deep regions of the target medium non-invasively.

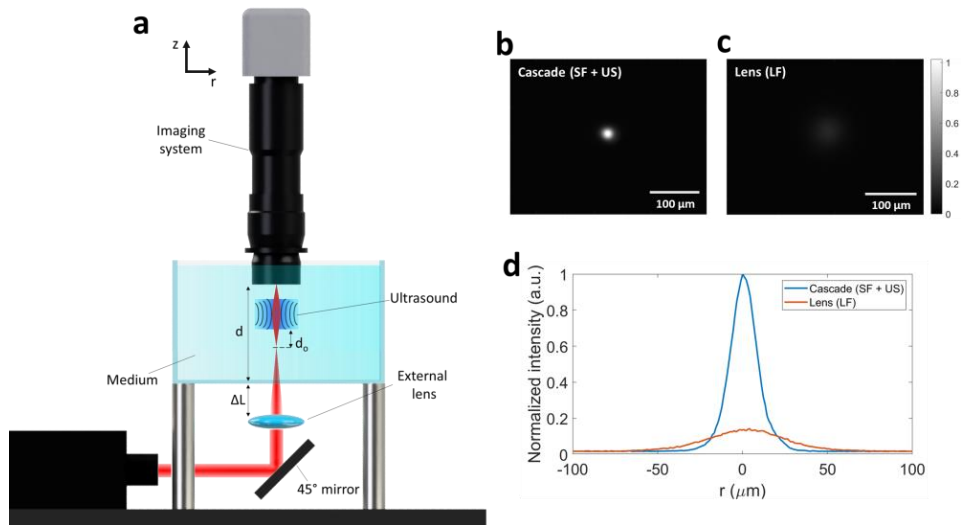

**FIG. 3** a) The schematic of the experimental setup showing the cascade system, where a short focal distance (SF) external lens is used to confine light through the medium, and a virtual ultrasonically defined GRIN waveguide is used to non-invasively relay the focused beam of light to a depth $d$ into the medium. The distance between the external lens and the medium is $\Delta L$, and the distance between the focus of the focal plane of the external lens and the virtual GRIN waveguide is $d_o$. b) The confined beam of light at the depth $d$ = 47.1 mm into the medium (DI water) using an aspheric lens with an effective focal distance $f$ = 25 mm (in air) and the virtual GRIN waveguide formed at $d_o$ = 13.47 mm. c) The confined beam of light at the depth $d$ = 47.1 mm into the medium (DI water) using a single aspheric lens with a longer effective focal distance (LF) $f$ = 50 mm. d) Radial cross-sections of the beams for the single external lens and the cascade system. The FWHM of the spot size for a single external lens is 49.4 µm, while for the cascade system it is reduced to FWHM=17.6 µm. The acquisition parameters were kept the same for both cases.

One of the advantages of the virtual GRIN waveguide is that by changing the shape and location of the ultrasound pattern, we can reconfigure the relaying function. For example, by forming the virtual waveguide at different distances ($d_o$) from the focal plane of the external lens, the location and size of the resulting focal spot can be changed. To demonstrate this reconfigurability, we performed experiments under the same setting, as shown in Fig. 3 using a cascade system, composed of the external lens ($f_2$ = 25 mm) and the virtual GRIN waveguide. The distance $d_o$ was changed by forming the same radially varying ultrasonic interference pattern at different axial distances through the medium. The cross-sections of the focused beam at different axial distances of light are plotted in Fig. 4a. The beam spot size (FWHM) and the peak intensity as a function of the axial distance ($d_o$) are plotted in Fig.4b and 4c, respectively. As the distance $d_o$ is increased, the beam size starts to decrease, and the intensity of the focused and relayed beam of light is increased. To understand this trend, we should note that when the distance between the focal plane of the external lens and the virtual GRIN waveguide, which acts as a GRIN lens, is small, the relayed focal spot will be magnified through the virtual GRIN waveguide. For example, for $d_o = 5.47$ mm, the relayed focused beam spot size (FWHM = 29.8µm) is larger than the beam spot size of light focused using the external lens (FWHM = 27µm), whereas at $d_o = 7.03$ mm (interpolated from the experimental results in Fig. 4b), the relayed beam spot size matches the spot size of the external lens. By further increasing $d_o$, the relayed beam spot size is decreased until it asymptotically approaches the diffraction-limited beam spot size of the virtual GRIN waveguide (i.e., FWHM = 15µm), which was directly measured using the same collimated input beam. This limit is shown as a dashed horizontal line in Fig. 4b. The peak intensity of light increases as the beam spot size is decreased until its rate of increase starts to slow down after $d_o = 17.47$ mm. There are two competing effects that contribute to the change in output intensity. On the one hand, the beam spot size is decreased, and on the other hand, the coupling efficiency to the virtual GRIN waveguide is decreased as $d_o$ is increased. This is mainly due to the loss of some photon flux coupled to the waveguide as $d_o$ is increased, since some photons will not be captured within the acceptance angle of the GRIN waveguide beyond a certain $d_o$, due to the divergence of the input beam. After this point, while the decrease in the focused beam size tends to enhance the intensity of the confined beam, the loss of photon flux tends to decrease the intensity of light. The competition between these two effects results in a much slower increase in the intensity of light after $d_o > 17.47$ mm, as shown in Fig. 4c. We expect the effect of losing photon flux becomes dominant compared to the decrease in the spot size for larger distances $d_o$, resulting in decreasing light intensity.

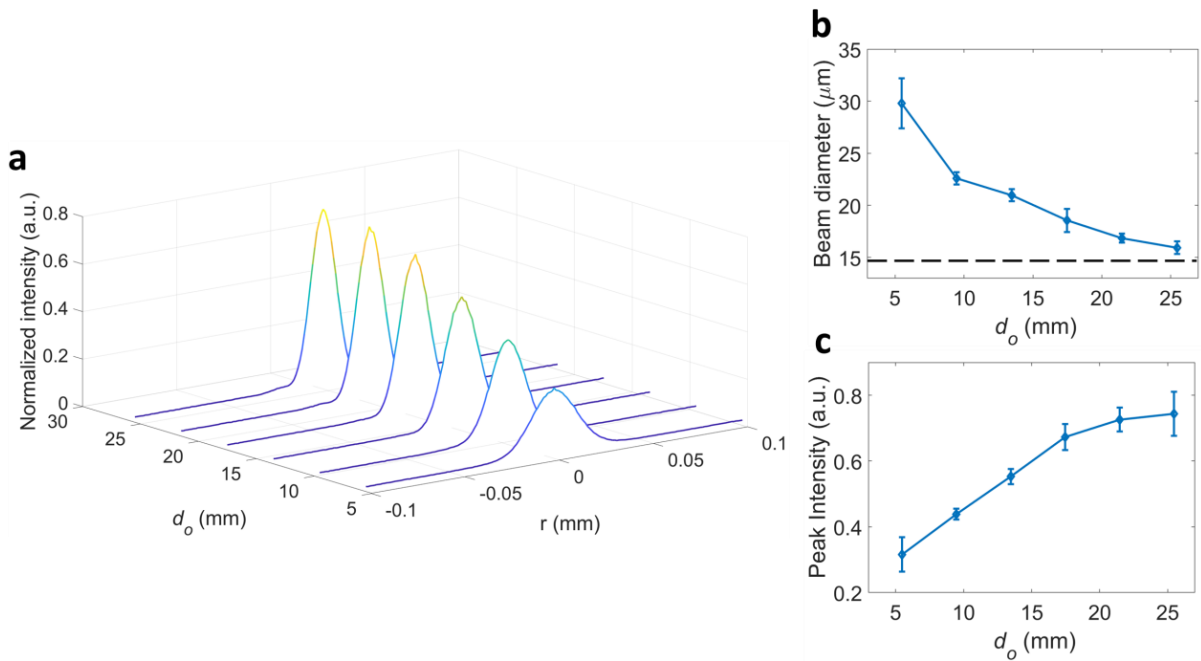

**FIG. 4** a) The radial cross-section of the confined beam of light through the cascade system of a short focal distance external lens with $f = 25$ mm (in air) and the virtual GRIN waveguide for different distances ($d_o$) between the focal plane of the external lens and the virtual GRIN waveguide. As the distance $d_o$ is increased, the spot size is decreased, and the intensity of light is increased. b) The spot size decreases as the distance $d_o$ is increased until it asymptotically approaches the diffraction-limited spot size of the GRIN waveguide shown by the dashed line. c) The peak intensity as a function of the distance $d_o$. The peak intensity increases as $d_o$ is increased until its rate of increase starts to slow down beyond $d_o = 17.47$ mm. The error bars represent the standard deviations of the measured values in different (n=4) experiments.

We also performed experiments in a scattering medium composed of Intralipid 20% mixed with DI water to compare the performance of the single external lens with a long focal distance (LF) and the cascade system composed of the shorter focal distance (SF) external lens and the virtual GRIN waveguide in a turbid medium. An external lens with a focal distance of $f_2 = 25$ mm (in air) in tandem with the virtual GRIN waveguide formed by ultrasound waves at the frequency of 1.512 MHz under the same conditions for experiments in DI water was used to focus light at a depth of $d = 47.1$ mm through the scattering medium. The distance between the focal plane of the external lens and the ultrasonically defined GRIN waveguide was set to $d_o = 13.47$ mm. The reduced scattering coefficient of Intralipid solution was measured as $\mu'_s = 0.93$ cm$^{-1}$ using the Oblique Incidence Reflectometry (OIR) method[18]. Therefore, light is confined through the depth of the scattering medium with an optical thickness of (OT = 4.38 TMFP). The confined beam of light at $d = 47.1$ mm is shown in Fig. 5a. We also performed an experiment using a single long focal distance $f_1 = 50$ mm (in air) external lens to directly focus light through the scattering medium at the same depth of $d = 47.1$ mm (Fig. 5b). The radial cross-sections of the focused beams are plotted for both cases of the single external lens with a long focal distance (LF) and the cascade system (SF+US) in Fig. 5c. The contrast ratio (i.e., the peak to background intensity ratio) for the case of the cascade system is measured to be 2.86, which is more than two times larger than the contrast ratio for the case of the single external lens with a contrast of 1.31. This shows that using the ultrasonic virtual waveguide in the cascade system, we can confine and focus light more effectively through the scattering medium, whereas the multiple scattering events in the medium almost overwhelms the beam of light focused by the single LF external lens. Also, as shown in Fig. 5c, the central peak of the beam focused by the cascade system is much narrower than that of the beam focused by the LF single external lens, showing that the same input light can be confined through the scattering medium with a much higher spatial resolution using the virtual ultrasonic GRIN waveguide.

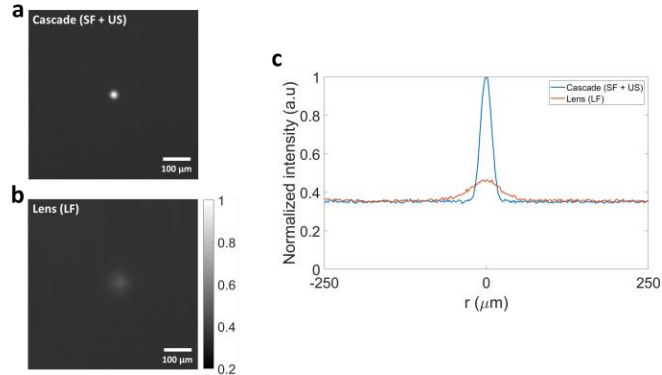

**FIG. 5** a) The confined beam profile of light at the optical depth of 4.38 TMFP into the scattering medium using a cascade system of an aspheric lens with a short focal distance (SF) $f = 25$ mm (measured in air) and the virtual GRIN waveguide formed at $d_o = 13.47$ mm. b) The confined beam of light using a single aspheric lens with a longer focal distance (LF) $f = 50$ mm (measured in air). c) Radial cross-sections of the beams for the single external lens and the cascade system.

We also performed another set of experiments in a medium made of a higher concentration of Intralipid with a slightly higher reduced scattering coefficient of $\mu'_s = 1.05$ cm$^{-1}$. The optical beam profiles through the medium with an optical thickness of OT = 4.95 TMFP are shown in Fig. 6a and 6b. The single external lens (LF) cannot focus light to the same depth anymore (Fig. 6a) due to the overwhelming effects of scattering in the medium, while the cascade system can still focus light through the medium (Fig. 6b). The radial cross-sections of the beams are plotted in Fig. 6c, where a contrast ratio of 1.6 can be achieved for the cascade system. These results demonstrate the power of the ultrasonic virtual waveguide to effectively confine and relay light through a thick scattering medium with higher contrast and spatial resolution compared to the single external lens.

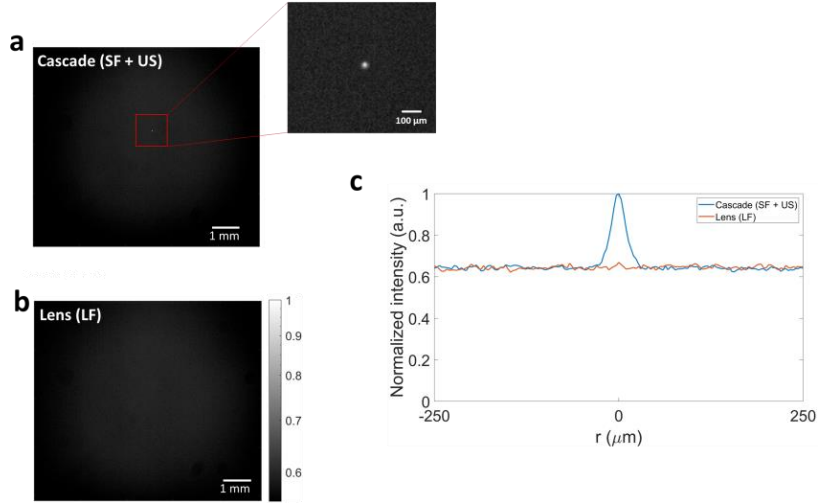

**FIG. 6** a) The confined beam profile of light at the optical depth of 4.95 TMFP into the scattering medium using a cascade system of an external aspheric lens with a short focal distance (SF) $f = 25$ mm (measured in air) and the virtual GRIN waveguide formed at $d_o = 13.47$. b) The confined beam of light using a single external aspheric lens with a longer focal distance (LF) $f = 50$ mm (measured in air). c) Radial cross-sections of the beams for the single external lens and the cascade system.

**Discussion**

In this paper, we showed that ultrasonically sculpted virtual GRIN waveguides can be formed deep in a medium non-invasively for in situ relaying and manipulation of light. In particular, we demonstrated that when the virtual waveguide is used in tandem with an external lens, the focal distance can be extended without sacrificing the spot size. In this cascade system of a short focal distance external lens and a virtual ultrasonically formed GRIN waveguide, the tightly focused beam of light by the external lens at shallow depth is relayed to a deeper region within the medium. We should note that a similar relaying effect could be achieved by placing a second physical lens, which can be a bulk GRIN lens, into the medium to extend the focal distance of the whole optical system without compromising the spot size. However, this would involve disrupting the medium by placing a physical optical component.

More complex optical systems, including compound lenses made of a cascade of physical elements have been designed for microscopy and photography[19,20]. Inspired from these established designs, the method presented in this paper based on ultrasonically formed virtual waveguides can be extended to design even more complex optical systems composed of multiple virtual optical waveguides and external physical elements to shape light within the target medium. A unique advantage of our ultrasonically sculpted optical waveguide is that it can be formed inside the target medium where a physical waveguide or a GRIN lens should not be inserted invasively.

An additional advantage of the presented technique is reconfigurability that opens up new opportunities for in situ non-invasive beam steering and beam forming inside the target medium. We demonstrated that the size and the peak intensity of the focused beam of light can be tuned by reconfiguring the pattern of ultrasound. This idea can be extended, for example, by combining the cascading method presented in this paper with our former work on spatial beam forming[14] to split a single externally focused beam of light into multiple focused beams of light deep inside the medium.

The comparison with a single long focal distance external lens to focus light through the medium at the same depth as the cascade system showed that the achieved level of confinement and the peak intensity of the focused beam are much higher when the cascade optical system is used. To ensure a fair comparison, the input beam diameters were kept the same. If the input beam diameter is increased, the spot size is decreased both for the cascade system and the single external lens at the expense of increasing aberrations. Of course, external physical lenses have been optimized (e.g., aspherical lenses) to minimize aberrations. In the case of the virtual ultrasonically sculpted waveguide discussed in this paper, the aberrations might be more pronounced when the input beam diameter is increased. These aberrations can be compensated by using adaptive optical techniques[21] or by optimizing the ultrasound pattern to sculpt a refractive index profile that minimizes spherical aberrations. Moreover, for specific applications, only a certain range of input beam diameters can be practical. For example, if an array of closely spaced optical channels is used to access the medium (e.g., for mapping or imaging), the size of the input beam for each channel needs to be kept small enough to enable dense integration without interference. For a given input beam diameter, the virtual GRIN waveguide needs to be formed properly such that the coupling of light from the external lens to the GRIN waveguide is optimized. An important advantage of the cascade optical system is that the output beam size and intensity can be tuned by changing the ultrasound pattern, independently of the input beam size, whereas for the single external lens with long focal distance, the output beam size and peak intensity only depend on the input beam diameter.

We demonstrated that the presented technique can also be used in scattering media to effectively confine and focus light at depth. We showed that using the cascade optical system, light can be confined and focused through a rather thick scattering medium (OT > 4 TMFP) with a noticeable contrast ratio of 2.86 between peak intensity and background. Moreover, we showed that the virtual GRIN waveguide can enable confinement of light through a scattering medium (OT = 4.95 TMFP), where the long focal distance external lens fails to confine light due to the overwhelming effect of scattering. Understanding the specific mechanisms that contribute to the observed enhancement of the contrast ratio for the light confined in turbid media using the cascade optical system needs further investigation. One of the possible contributing factors can be the geometrical path difference between the cascade optical system and the long focal distance external lens, which affects the number and distribution of both ballistic and scattered photons that reach the focal point and the surrounding regions. Also, some of the scattered photons can be guided and confined through the virtual GRIN waveguide towards the focal point, potentially contributing to the enhancement of the overall light throughput at the target location.

The observed unprecedented advantage over conventional external optics for non-invasive manipulation of light in the target medium can enable a plethora of applications involving light delivery and potentially optical imaging, where photons should be collected and relayed through the medium.